\documentclass[a4paper,11pt]{article}
\pdfoutput=1 

\usepackage{jinstpub} 
\usepackage{wrapfig}
\usepackage{caption}
\usepackage{subfig}

\title{\boldmath Linearity and rate capability measurements
of RPC with semi-insulating crystalline electrodes
operating in avalanche mode}

\author[a]{A. Rocchi\note{Corresponding author.}}
\author[a]{, R. Cardarelli}
\author[a]{, B. Liberti}
\author[b]{, G. Aielli}
\author[b]{, E. Alunno Camelia}
\author[b]{, P. Camarri}
\author[b]{, M. Cirillo}
\author[a, b]{, A. Di Ciaccio}
\author[b]{, L. Di Stante}
\author[b]{, M. Lucci}
\author[a]{, E. Pastori}
\author[a]{, L. Pizzimento}
\author[b]{, G. Proto}
\author[b]{, E. Tusi}

\author[b]{, and R. Santonico}

\affiliation[a]{INFN Roma Tor Vergata,\\Via della ricerca scientifica 1, 00133 Roma, Italia}
\affiliation[b]{Dipartimento di Fisica, Universit\'a di 
Roma Tor Vergata,\\Via della ricerca scientifica 1, 00133 Roma, Italia}

\emailAdd{alessandro.rocchi@roma2.infn.it}

\abstract{The intrinsic rate capability and the ageing properties of the Resistive Plate Chambers are closely related to the electrodes material and to the front-end electronics threshold. The development of a low noise pre-amplifier led us to improve the intrinsic rate capability of High Pressure Laminate (bakelite) up to  $\sim10\;kHz/cm^2$, nevertheless the effective rate is significantly limited by electrodes ageing. To further improve the effective rate capability new materials are investigated. A Resistive Plate Chamber with crystalline semi-insulating Gallium Arsenide electrodes has been characterized with high energy electrons beam at the Beam Test Facility (BFT), (INFN National Laboratory of Frascati, Italy). The response of the Resistive Plate Chamber to multiple bunched electrons was measured operating the detector in avalanche mode. The intrinsic rate capability has been also measured operating the detector in a uniform high energy gamma radiation field at the GIF++ facility (EHN1 of SPS, CERN). }


\arxivnumber{1234.56789} 

\proceeding{XV$^{\text{th}}$ Workshop on Resistive Plate Chambers and Related Detectors\\
  10-14 February 2020\\
  Rome}

\begin{document}
\maketitle
\flushbottom

\section{Introduction}
\label{sec:intro}
RPCs detectors are widely used in high energy physics experiment because of their excellent intrinsic time resolution and response. The concept behind the detection mechanism allows producing high strength and cost effective detectors \cite{Sant}. For this reason this kind of detector has been used mainly to cover large area experiments. 
The electronics front-end upgrade and the progress in the gas gap design have improved the RPCs rate capability performances up to $10\;kHz/cm^2$ \cite{Atlas}. Nevertheless, the total charge, integrated during the detector operations, produces an increase of the electrode bulk resistivity resulting in a significant rate capability degradation \cite{ageing}. For this reason, the RPCs detectors should work at a fraction of their intrinsic rate capability depending on the experiment conditions and lifetime. 
It can be concluded that the effective rate capability of the RPC detectors is limited mainly due to the electrodes ageing problem and that new materials should be investigated.

 The RPCs signal charge doesn't show a sharp distribution when the detector is operated in a low gain avalanche mode. The detector response is not proportional to the energy released by a single detected particle. If the number of synchronous interactive particles becomes high, the central limit theorem can be applied and the signal average charge is proportional to the number of bunched particles. This feature was exploited in the past to measure extensive air showers, paving the way of calorimetry to the RPC detectors \cite{Iacovacci}. In the recent years many proposal were moved to use RPC as the active area of the calorimeter in both cosmic ray and collider experiments \cite{Iuppa} \cite{DHCAL}. The RPC linear limit was measured only for the streamer operating mode. In the saturated avalanche mode, the signal charge shows large fluctuations with respect to the average value, but the linear limit should increase proportionally to the average charge reduction. RPC operated in saturated avalanche mode, therefore, could be exploited in that situations where the synchronous particles density is very high. 

\section{Detector design}
\begin{figure}[htp]
\centering
\hfill
\begin{minipage}[b]{.45\columnwidth}
  \centering
  \includegraphics[width=6.cm]{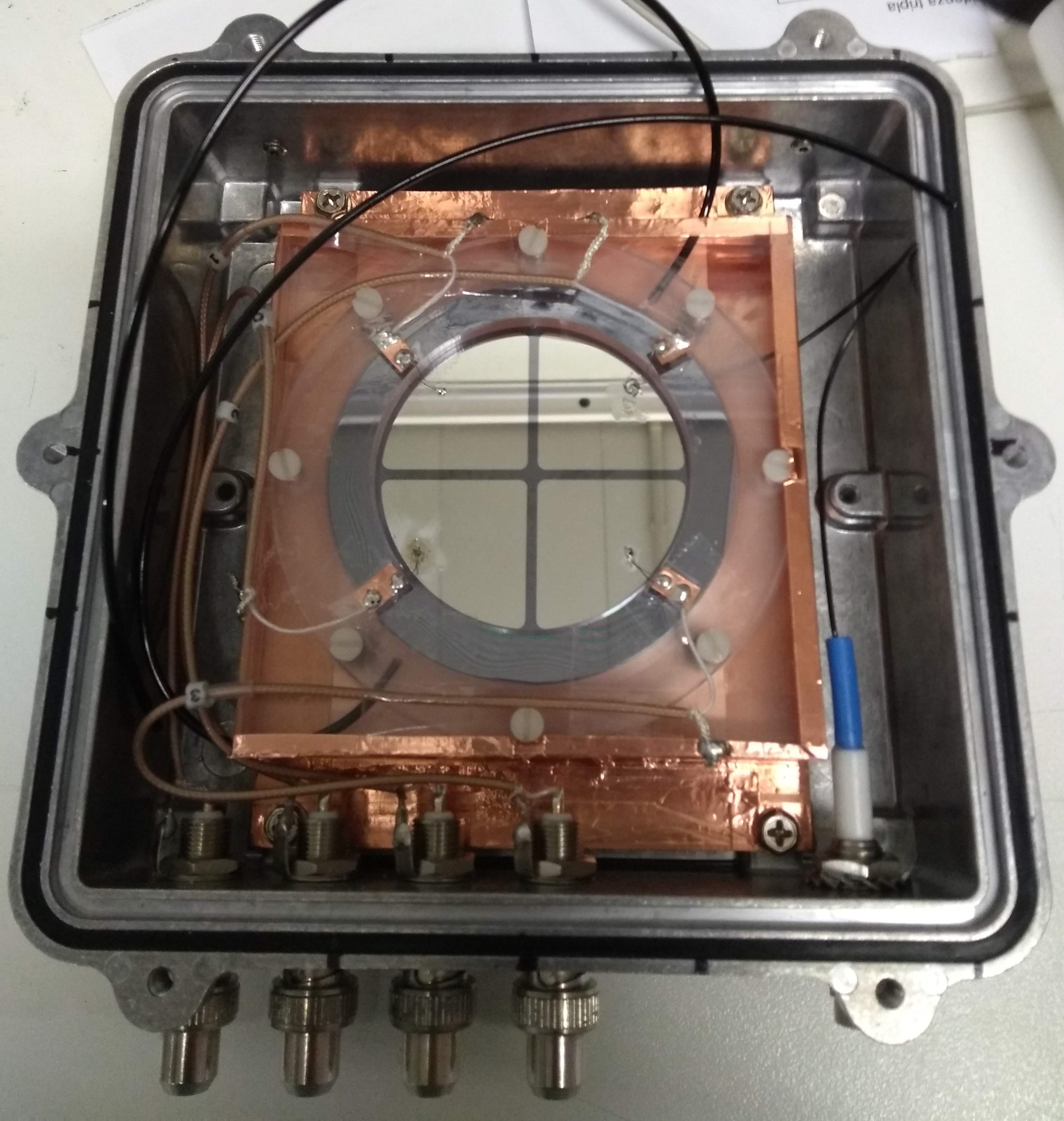}\quad
  \caption{\small{\emph{A picture of the detector hosted in the Aluminium box with all the services connected.}}}
  \label{Prot2_pic}
\end{minipage}\hfill
\begin{minipage}[b]{.45\columnwidth}
  \centering
  \begin{tabular}{|l|l|l|}
  \hline                                                            
  \textbf{\small{Material}}	&\small{SI undoped GaAs}\\[3pt]
  \textbf{\small{Thickness}}	&\small{$640-643;\mu m$}\\[3pt]
  \textbf{\small{Diameter}}	&\small{$3''$}\\[3pt]
  \textbf{\small{Resistivity}}	&\small{$1.4\times10^8\;\Omega cm$}\\[3pt]
  \textbf{\small{Surface treatment}}	&\small{both polished}\\[3pt]
  \textbf{\small{Growth method}}	&\small{VGF}\\[3pt]
  \textbf{\small{Orientation}}	&\small{$(100)\pm0.01^\circ$}\\[3pt]
  \textbf{\small{Mobility}}	&\small{$5300\;cm^2$/Vs}\\[3pt]
  \textbf{\small{EPD}}	&\small{$2016\;/cm^2$}\\[3pt]
  \textbf{\small{Nc}}	&\small{$8\times10^6/cc$}\\[3pt]
  \hline
  \end{tabular}
  \captionof{table}{\small{\emph{Semi-insulating Gallium Arsenide wafers specifications.}}}\label{tab:GaAs2_spec}
\end{minipage}\hspace*{\fill}
\end{figure}

In this paper the results obtained through RPCs with semi-insulating Gallium Arsenide (SI-GaAs) electrodes, designed for high energy physics, are described. The electrodes were produced by ITME  \cite{ITME} and all the technical specifications are listed in table \ref{tab:GaAs2_spec}. As compared to the high pressure laminate, the SI-GaAs has a crystalline structure with a high carriers mobility, so that the bulk conductance is limited only by the very low free carriers concentration. The SI-GaAs bulk resistivity, two orders of magnitude lower then that of high pressure laminate, leads to better rate capability performances.  
The electrodes holder consists of three parts: a spacer and two frames. The spacer hosts the gas inlet, the gas outlet and the electrodes, keeping them $1\;mm$ spaced. The frames push the electrodes  on the spacer. The holder dimensions are $10\;cm\times10\;cm\times0.8\;cm$ and the detector active area is $25\;cm^2$. The device is hosted in a watertight Aluminium box that shields it from electromagnetic noise.\\
The electrodes contacts are made by sputtering Aluminium on the wafers surface and consist in four pads on both the high voltage and ground side. Each pad is $\sim6.25 cm^2$ large. The pads bonding is made with a silver paint drop. On the high voltage side the pads are connected in parallel; on the ground side, instead, each pad is connected to the ground plate through $100\;k\Omega$ resistor. 
A picture of the detector is shown in figure \ref{Prot2_pic}.

\section{Linearity measurements at BTF (LNF)}
\begin{figure}
\centering
\hfill
\begin{minipage}[b]{.4\columnwidth}
\centering
\includegraphics[width=5cm]{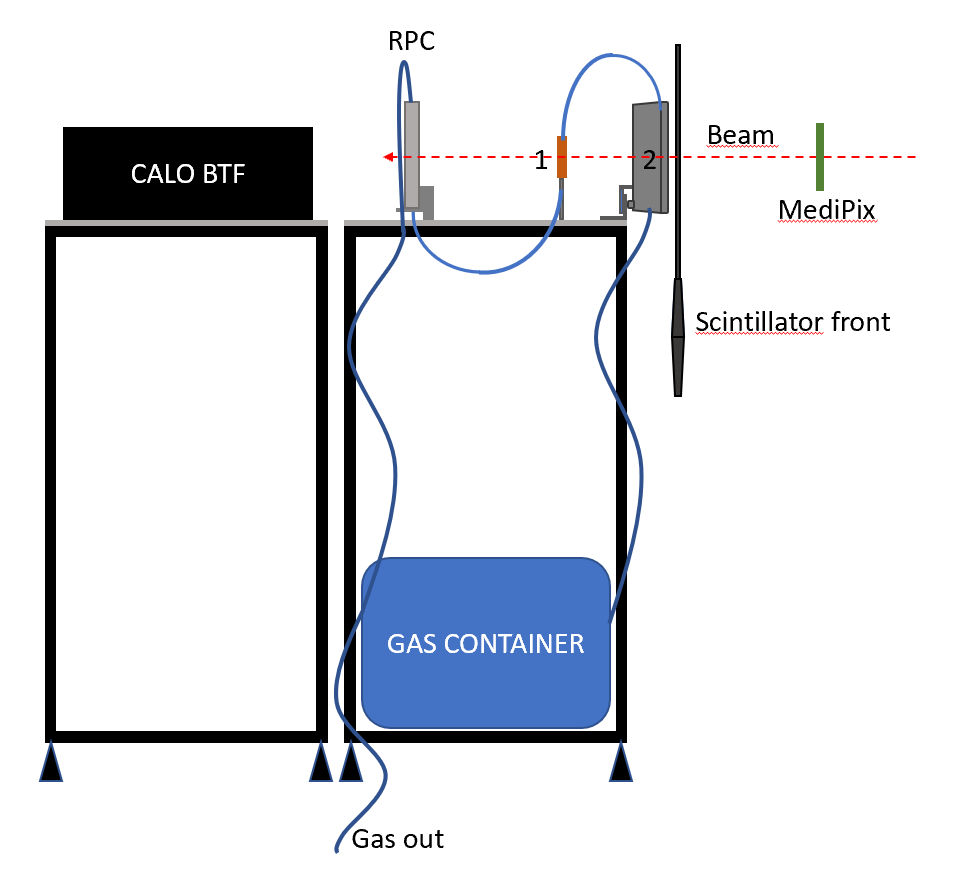}
\caption{\small{\emph{Mechanical frame for the BTF experimental test.}}}
\label{btf_setup}
\end{minipage}\hfill
\begin{minipage}[b]{.5\columnwidth}
\centering
\includegraphics[width=8.5cm]{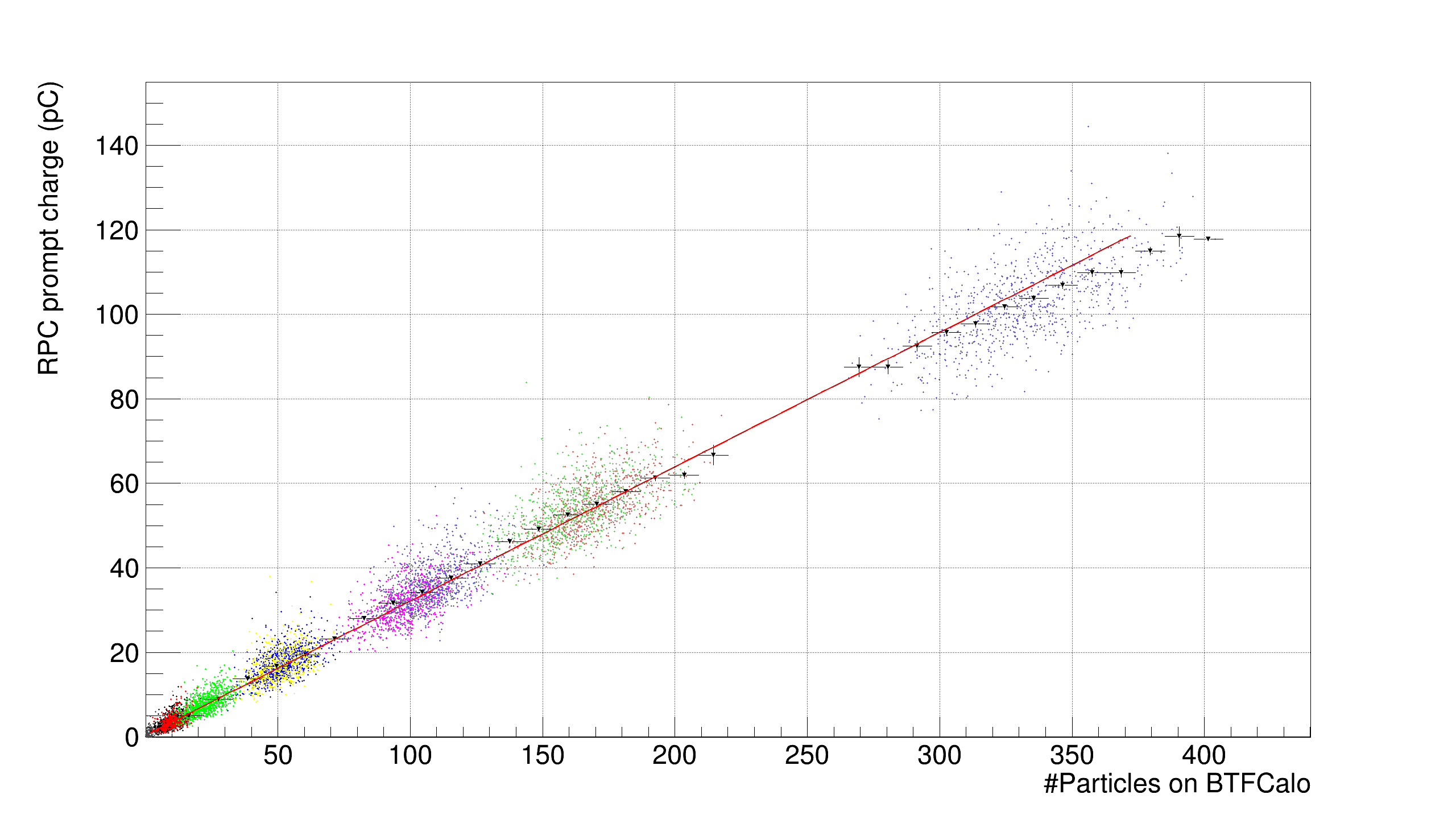}
\caption{\small{\emph{Prompt charge of the RPC operated in avalanche mode as function of the particles multiplicity measured with the calorimeter.}}}
\label{Linearity tot}
\end{minipage}\hspace*{\fill}
\end{figure}

This linearity study is based on the tests performed by M. Iacovacci and S. Mastroianni at the BTF of Laboratori Nazionali di Frascati, in which the intrinsic linearity of bakelite RPCs operated in streamer mode has been measured at different impinging particle densities \cite{Iacovacci}. The upper limit for the linearity response, is expected to increase of at least a order of magnitude with the streamer to avalanche transition due to the average charge reduction.  
The linearity response was studied using secondary electrons beam with energy of $250\;MeV$ produced at the BTF of Laboratori Nazionali di Frascati \cite{BTF}. The particles bunch multiplicity was changed in a range between $1\;particle/bunch$ to $\sim300\;particles/bunch$. The mechanical frame holding the detectors is shown in figure \ref{btf_setup}. 
The detector used for this measurement is tagged with the number two. Detectors provided from the facility were used to control the beam properties: the Medipix detector worked as real time monitor for the beam position and intensity, the BTF lead glass Cherenkov calorimeter measures the beam intensity. The premixed gas mixture consists of $95\%$ C$_{2}$H$_{2}$F$_{4} + 4.5\%$ iC$_{4}$H$_{10} + 0.5\%$  SF$_{6}$  and was stocked in a small $8\;l$ container placed at the base of the trolley. The ATLAS-like RPC detector worked as reference to monitor the gas conditions. The frontal plastic scintillator signals were acquired to improve the geometric acceptance during off-line analysis. The total interacting material in front of the calorimeter consists of $\sim0.1$ radiation length. The beam profile had a Gaussian shape with a $0.8\;mm$ $\sigma_y$ on the vertical direction and a $2.5\;mm$ $\sigma_x$ on the horizontal direction. In the test reported in \cite{Iacovacci}, the beam spot had the same dimensions as the vacuum pipe ($5\times3\;cm^2$) and was entirely unfocused. In the presented test, this configuration was not usable because of the detector dimensions, whose pad is a quarter-circle with radius $2.8\;cm$. The bunch frequency was $\sim20\;Hz$ and the bunch time width was $10\;ns$: it is assumed that the electrode material, although it is a main structural detector parameter,  has a marginal role in processes that occur within a sub-microsecond time scale. The detector effective high voltage was set to $6100\;V$ where the detector efficiency, without front-end amplifier, is $85$\% and the average induced prompt charge is $0.45\;pC$. 
The results of the intensity scan were presented in figure \ref{Linearity tot}. The RPC prompt charge was plotted as function of the multiplicity measured with the calorimeter. Different colors correspond to different acquisitions.

\begin{figure}[h]
\centering
\includegraphics[width=15cm]{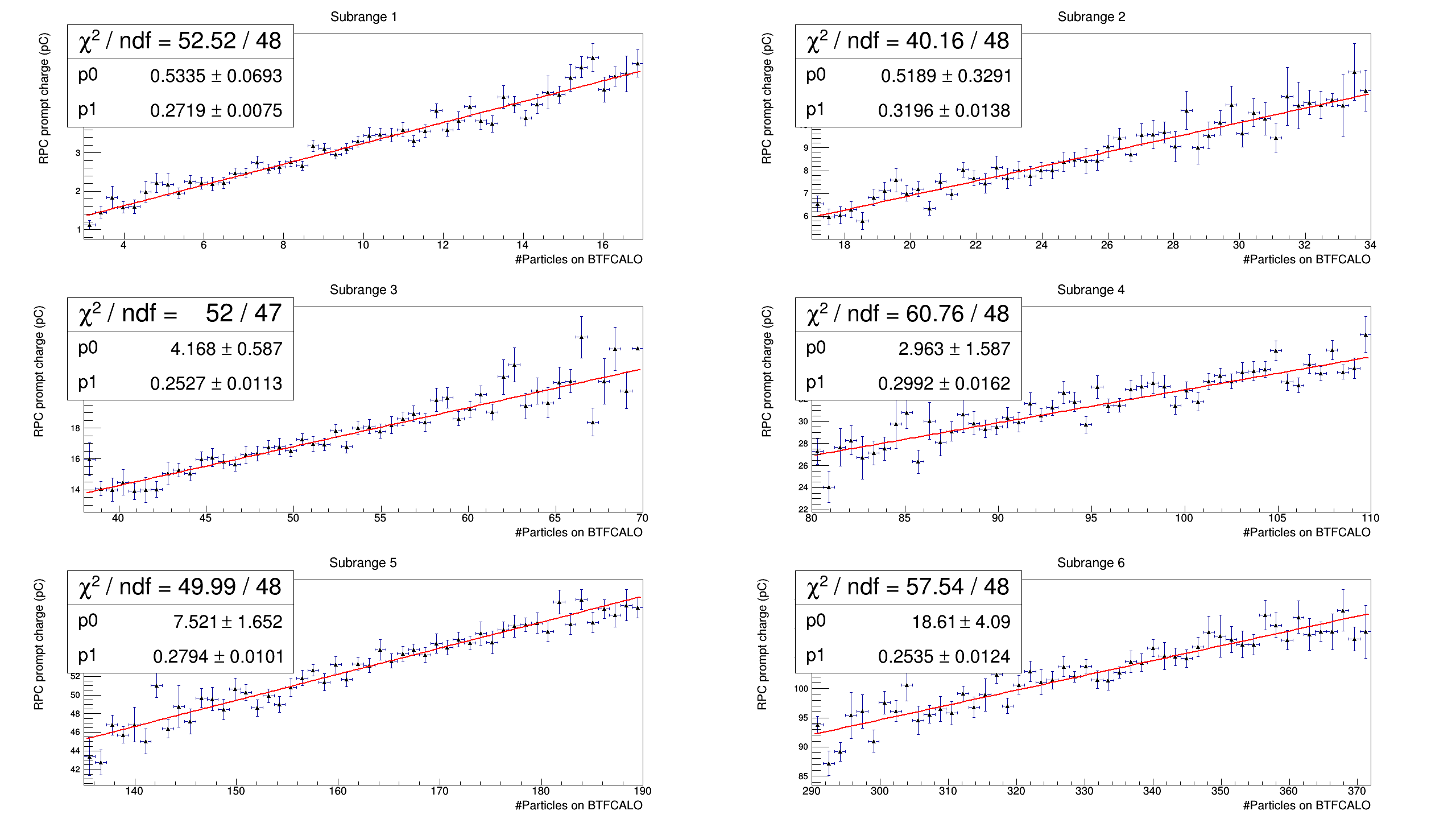}
\caption{\small{\emph{ Linear fits for different sub-ranges.}}}
\label{linearity_fit}
\end{figure}

The statistical analysis of the data set was performed dividing the whole interval in six sub-ranges centred with respect to the average intensity values set during the scan. In order to locally check the linearity response, a linear fit on the RPC average charge has been performed for each sub-range. The fit results are shown in figure \ref{linearity_fit}. The p-value calculated with the $\chi^2$ of the linear fits, for each sub-range, results smaller than $0.05$ which is the lower limit for the $95$\% confidence level, therefore the linear model is consistent with the data sets. As the last step of this statistical analysis, the agreement of the fit residuals, defined as the normalized difference between the observed and the fit value, with the Gaussian distribution was verified. 

The slope extrapolated by the linear fits for each sub-range does not show a relation with the particles multiplicity . The intercept, instead, increases systematically with the sub-range index. This result could be caused by systematic errors related to small variations of the experimental parameters neglected during the data acquisition as the reference calorimeter high voltage and the beam position fluctuations. The Gaussian shape of the beam profile does not allow defining a constant particles density. For this purpose a lower limit for the particle density is defined approximating the Gauss distribution with a uniform distribution within $2\; \sigma$ range centred on the beam center.  
The area of the considered region is enclosed by an ellipse with major and minor semi-axis $\sigma_x$ and $\sigma_y$ respectively, and measurements $\sim 6.28\;mm^2$.  An electron has $\sim47$\% chance of being in this region, therefore, for $300\; particles/bunch$ multiplicity, the particles density is  $\sim 22\times 10^6\;particles/m^2$. 
It can be concluded that the RPC detector, operated in avalanche mode, has a linear response in each measured multiplicity subinterval with a confidence level greater than 95\% up to $\sim22\times 10^6\;particles/m^2$. 



\section{Rate Capability measurement at GIF++ (EHN1)}

\begin{figure}[h!]
\centering
\includegraphics[width=10cm]{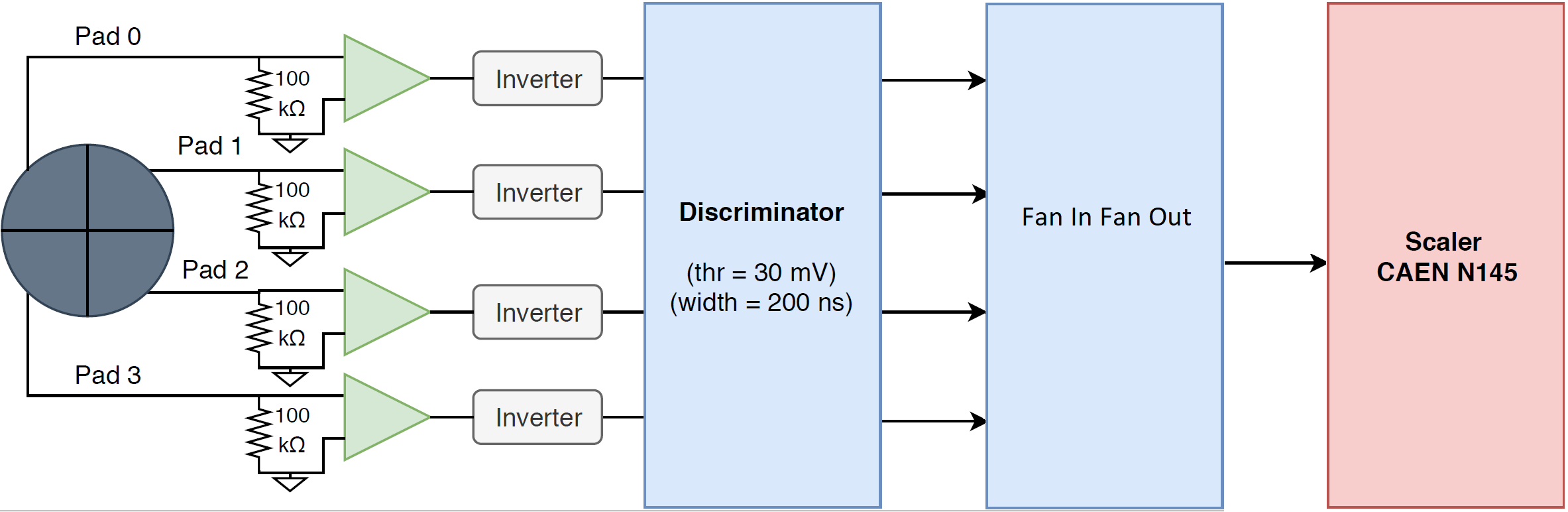}
\caption{\small{\emph{Experimental set-up for the counting rate measurement: the signals were amplified, inverted and discriminated. The signals logic OR rate was measured with a scaler.}}}
\label{Rate_cap}
\end{figure}

The rate capability measurement was performed at the Gamma Irradiation Facility located at the Experimental Hall North 1 of CERN. The radiation field was produced by a $14.9\;TBq$ $^{137}$Cs source and could be attenuated through absorption filters with different values. The test was performed by placing the detector at $1.5\;m$ in front of the source on the downstream side and measuring the total current and the counting rate by varying the absorber filters. An Atlas like RPC was placed beside the detector under test in way to compare the current response. The readout scheme for the counting rate measurement is described in figure \ref{Rate_cap}. The signals from the four pads were amplified with a low noise front-end charge amplifier \cite{Cardarelli:FE}, discriminated, processed with a logic or, and counted with a scaler in a $50\;s$ time interval. The discriminators threshold was set to $\sim 17\;fC$ and the NIM signals shape was formed with $200\;ns$ time width.

The measurement was repeated for three different values of the absorber. The facility simulation \citep{gif_irad} show a maximum photons current of $\sim10^7\;cm^{-2}s{-1}$ with the filters in position $1$ (no absorber), therefore, assuming the RPC gamma conversion factor is $5\times10^{-3}$, the maximum detectable rate was $5\times10^4\; cm^{-2}s{-1}$. The current measured without absorber was divided by that measured for different absorption factors. The relation \ref{eq_abs1} shows how, at fixed average charge $\bar{Q}$, the ratio between the currents $I$, measured with different absorption factors, should be consistent with the ratio between the absorption factors. The deviation of the current ratios from the theoretical absorption factor ratios indicates that the counting rate saturates to a value lower than the effective particle flux $\Phi$. The same considerations hold for the ratios between counting rates measured at different absorption factors.   

\begin{equation}
\frac{I_{ABS1}}{I_{ABSn}}=\frac{\Phi_{ABS1}\bar{Q}}{\Phi_{ABSn}\bar{Q}}=\frac{n\Phi_{ABSn}}{\Phi_{ABSn}}=n
\label{eq_abs1}
\end{equation}

\begin{figure}
\centering
\hfill
\begin{minipage}[b]{.47\columnwidth}
\centering
\includegraphics[width=7.3cm]{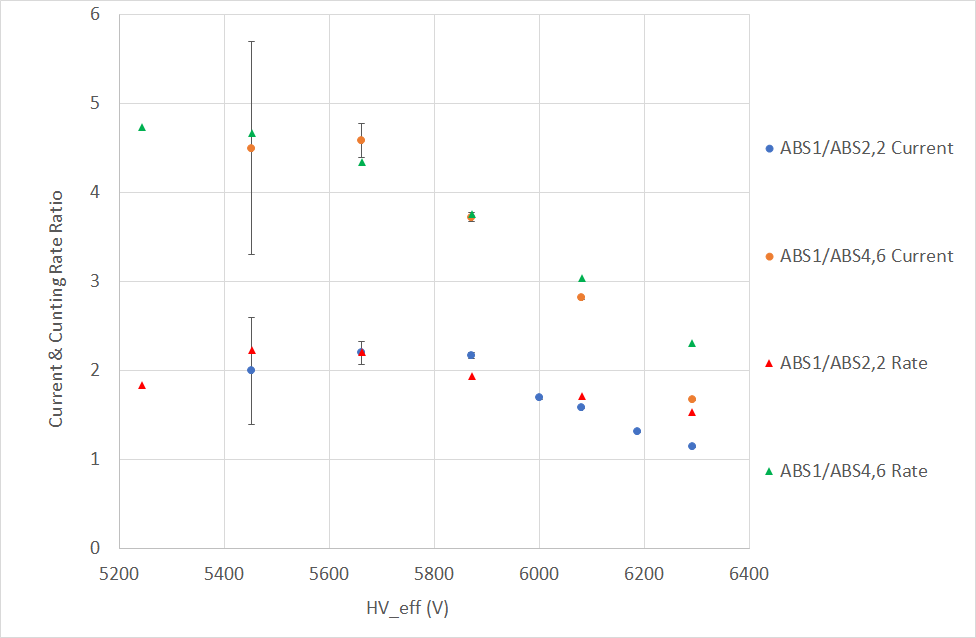}
\caption{\small{\emph{The counting rate ratios (triangles) and the current ratios (circles) as function of the high voltage working value.}}}
\label{Rate1}
\end{minipage}\hfill
\begin{minipage}[b]{.48\columnwidth}
\centering
\includegraphics[width=7.5cm]{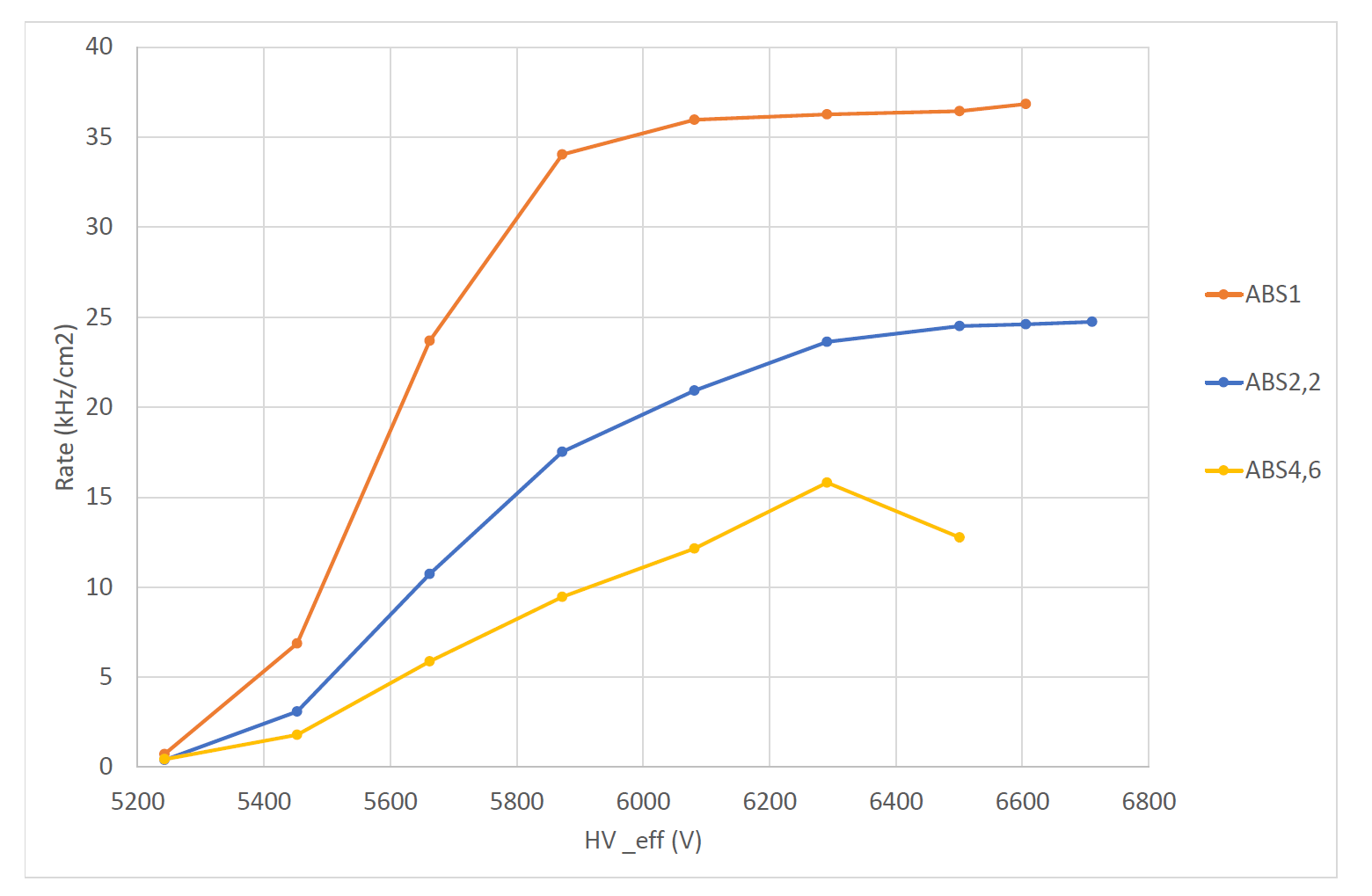}
\caption{\small{\emph{Measured counting rate for three different value of the absorption factor as a function of the high voltage.}}}
\label{Rate2}
\end{minipage}\hspace*{\fill}
\end{figure}

The current ratio (circles) and the counting rate ratios (triangles) as a function of the high voltage working point are shown in figure \ref{Rate1}. The detector under test, shows a good response for the first three high voltage working points, where current and counting rate ratios are consistent with the absorption factor ratios. 
 The counting rate measurement results are shown in figure \ref{Rate2}: the maximum measured counting rate is $36\times 10^3\;cm^{-2}s^{-1}$. The rate value measured at $5870\;V$ without absorber is $34\times10^3\;cm^{-2}s^{-1}$. With the source absorber set to $2.2$ the counting rate results to be $17.5\times 10^3\;cm^{-2}s^{-1}$ which gives $\sim10\%$ saturation. With the source absorber set to $4.6$ the counting rate results to be $9.5 \times 10^3\;cm^{-2}s^{-1}$ which corresponds to a saturation of $\sim22\%$.

The developed detector is capable of counting up to $\sim30\times 10^3\;cm^{-2}s{-1}$ photons. This value is significantly limited by the high threshold value set to $\sim 17\;fC$, nevertheless the result is more than four times higher than the measured rate capability of $1\;mm$ gas gap HPL electrodes RPCs with $\sim3\;fC$ threshold \cite{Atlas}.
The saturation effect must be further investigated performing more accurate measurements.

\section{Conclusions}
RPCs with Semi-Insulating electrodes have been found simple to build and cheap. The excellent wafer surface uniformity allows using this material without linseed oil coating. The linearity limit of the RPC operated in avalanche mode was studied and it can be claimed, with a confidence level above $95\%$, that the RPC response is linear in each considered sub-range up to $\sim 22\times 10^6\;particles/m^2$. 
The maximum measured counting rate is $\sim37\;kHz/cm^2$ when the detector works in a uniform gamma radiation field, whose flux is $\sim10^4\;kHz/cm^2$. The ratios between the currents and the counting rates , measured with different background absorption factors, agree with the expected values for working voltages up to $5700\;V$ and with the discrimination threshold set to $\sim17\;fC$. A saturation of $22\%$ was observed at working voltage set to $5870\;V$, where the detector efficiency was measured to be $85\%$. The discrimination threshold could be lowered down to $\sim3\;fC$ improving the electromagnetic shielding so that the efficiency knee point could be moved down to at least $5600\;V$, where no saturation effects were observed. 


\end{document}